# Bootstrap Nonlinear Regression Application in a Design of an Experiment Data for Fewer Sample Size

**Bello, Oyedele Adeshina[1]; Bamiduro, Timothy Adebayo[2]; Chuwkwu, Unna Angela [2]& Osowole, Oyedeji Isola[2]**

1. Department.of Statistics Federal University of Technology, Minna, Niger State, Nigeria.
oyedele.bello@futminna.edu.ng[1]
2. Department of Statistics University of Ibadan, Nigeria.
au.chukwu@ui.edu.ng[2]

**Abstract—**

*This paper reports on application of bootstrap nonlinear regression method to a design of an experiment dataset with fewer experimental runs. Design with desired properties was augmented and verified using graphical techniques. The augmented design with the desired properties benefited the accuracy of the approximated function used. The computation power of R-language and SAS for computing nonlinear function and bootstrap was also compared.*

## Introduction

The beauty of statistics is in its ability to obtain an approximation for the function $(f)$, that is able to describe a phenomenon of interest, but often time some phenomenon generate little or no data (insufficient information or sample size) which makes it difficult to obtain sufficient, efficient and robust estimates. Nevertheless, finding an approximate function to describe the relating variables in such phenomenon is still of interest. Therefore, we need to look for a way around the problem.

### 1.1 The Problem of Data Sufficiency and the Cost of An Experiment

The availability of sufficient information or sample size about a phenomenon is not always at the researcher luxury. Some phenomenon could hardly produce up to ten sample size, and the estimate gotten from such small observations could be misleading in application. How many of such fewer samples will the law of large numbers be applicable to? Phenomenon in discipline such as Medical Science, Geotechnical Engineering where often data were simulated to assume the phenomenon behaviour, while in experimental design we are faced with the problem of cost of experimenting and large number of experimental runs which sometime could be unattainable. Researcher can be interested in studying a relative new or rare phenomenon, which there is little or no prior knowledge about the phenomenon or data about the phenomenon, for example- information on a newly discovered virus spread, sud-den movement of the galaxies or earth crust in away it has never done before, the little jot of information from such newly discovered phenomenon can be mimic or better still simulated to generate a much larger datasets. But simulation can be highly prone to error because most of the simulation methods tell us little or nothing about the distribution of such simulated data, if at all it does, does the simulated data conform to the true distribution of the simulated phenomenon. This work shall prefer to re-sample such token information or sample to give birth to a much larger sample size using the bootstrap re-sampling method. The name "bootstrap" incidentally conveys the impression of "something for nothing" where statisticians idly re-sampling from their samples (F.W. Scholz, 2007). Bootstrap re sampling method cancelled out the problem of large variance which usually arises when modelling with small set of samples.

## 1.2 Data Visualization:

The first step to computing in statistics is to look at your data and ask researchable questions on it. The method of data visualization can aid in looking indept into the data so as to be able to ask necessary questions about the phenomenon of study. The failure of taking advantage of the exploratory data analyses (EDA) or any other method of data visualization usually results to using a correct answer to answer a wrong question. The method of data visualization helps to suggest the possible relationship existing between variables involved and in generality a close and relevant function can be approximate. According to Julian J. Faraway (2002) "Statistics starts with a problem, continues with the collection of data, proceeds with the data analysis and finishes with conclusions. It is a common mistake of inexperienced Statisticians to plunge into a complex analysis without paying attention to what the objectives are or even whether the data are appropriate for the proposed analysis. Look before you leap!" One important way to look before leaping is to visualize your data so you don't prescribe a linear function for a non-linear relationship.

## 1.2 Exploratory Data Analysis (EDA):

This method of analysis uses visualization tools and computes synthetic descriptors. EDA is required at the beginning of the statistical analysis of multidimensional data, in order to get an overview of the data, transform or recode some variables orient further analyses (Daniel Borcard et al, 2011). Data visualization helps to detect outliers, spotting local structures, systematic errors in the data, skewed or unusual distributions trends, clusters, and for evaluating modelling output and presenting results. Note, deciding on which graphics to use is often research and researchers view dependent. Most importantly, EDA is a major way of suspecting a nonlinear dataset in a regression problem.

## 1.3 Exactness VS. Approximate Function-The Avoidance Of Nonlinear Problems

Schabenberger and Pierce (2002) charge that researchers tend to avoid non-linear models because of discomfort with the messiness of implementation. Whereas the limitation of linear models that constrained in how the parameters and coefficients can interact is a big drawback to correct modelling of many biological, agro metric and many other real life scenarios that are usually nonlinear in nature.

In a linear model we have had to ensure that the normal equations, which are the first derivatives of the objective function with respect to the coefficients, were independent of the coefficients. This constraint has precluded us from using a wealth of biologically realistic model forms. The messiness is unavoidable; the avoidance of non-linear is basically because of its numerous numerical iterations and being highly computer-intensive in solving the problem.

However, it is arguably better to get an approximate answer to a meaningful question than to get an exact answer to an approximation to a meaningful question (Tukey, 1962). Despite the passive challenges to solving a nonlinear problem we should choose model forms that genuinely represent the phenomenon under study. (See Andrew P. Robinson and Jeff D. Hamann; Forest Analytics with R- An Introduction pg. 199-203 for details).

Advantages of Nonlinear Models

Nonlinear models are often derived on the basis of physical, chemical or biological considerations, also, from differential equations, and have justification within a quantitative conceptualization of the process of interest.

The parameters of a nonlinear model usually have direct interpretation in terms of the process under study.

Constraints can be built into a nonlinear model easily and are harder to enforce for linear models. Truly, fitting a nonlinear regression model to data is slightly cumbersome but we would usually prefer such a model whenever possible, rather than to its alternative, perhaps less realistic linear model.

## 2 METHODOLOGY

Considering;

$$y = f(x)\beta + \varepsilon \qquad (1)$$

Where $x = (x_{1,} x_{2,} ..., x_k)$, $\beta$s is the coefficient estimates, $\varepsilon$ (error term) ~ N (0; 1) and independent. $f$ is the function that describes the form in which the response and the input variable are related, its mathematical form is not known in practice. It is often approximated within the experimental region.

### 2.1 The Design

Factorial Design is a class of orthogonal design used for fitting first order model, this work makes use of factorial design with four levels. The central composite design was used to fit the second order. Fitting second order model also means augmentation of additional points to the initial first order design. The full factorial runs with na = 2k points on the axis of each factor at a distance from the center of the design.

The center runs n0(0,0...,0) without replication. The total number of runs N=nf + na + no.

The practical deployment of a CCD often arises through sequential experimentation, as that is, when a factorial design has been used to fit a first-order model, this model has exhibited lack of fit, and the axial runs are then added to allow the quadratic terms to be incorporated into the model. There are two parameters in the design that must be specified: the distance α of the axial runs from the design centre and the number of center points n0. See Appendix III for the constructed designs

### The Desired Properties

Orthogonality: A design D is said to be orthogonal if the matrix XX is diagonal, where X is the model matrix of $y = X\beta + \varepsilon$. The elements of $X'X$ will be uncorrelated because the off diagonal elements of $V(X'X)$ will be zero.

Rotatability: A design D is said to be rotatable if the prediction variance $Var[\hat{y}(x)] = \delta^2 (X'X)^{-1} f(x)$ is constant at all points that are equidistant from the design centre. The prediction variance remains unchanged under any rotation of the coordinate axes.

[1] Uniform Precision: A rotatable design is said to have the additional uniform precision property if Var [ y^ (x)] at the origin is equal to its value at a distance of one from the origin. This helps in producing some stability in the prediction variance in the vicinity of the design centre. For details on RSM, Design and properties See; Andre I Khuri et al (2010), Khuri, AI, Cornell (1996) , Bello O.A. (2014). "Modeling Cassava Yield: A Response Surface Methodology Approach.", Unpublished M.Sc Thesis, University of Ibadan Nigeria, and Bello O. A. ( 2014) International Journal on Computational Sciences & Applications (IJCSA) Vol.4, No.3, June 2014, 63 for details.

## 3 THE INVERSE POLYNOMIAL – A NONLINEAR FUNCTION

The Inverse Polynomial –A Nonlinear Function

The general forms of Inverse Polynomials of (Nelder, 1966; Nduka, 1994; Holger Dette, 2007) are family of non-linear function and it is intrinsically nonlinear.

$$\prod_{i=1}^{k} \frac{x_i}{q} = \prod_{i=1}^{k} (\beta_0 + \beta_{1i} x_i + \beta_{2i} x_i^2 + \ldots + \beta_{p-1i} x_i^{p-1}) \quad (2)$$

Where, k = number of factors q = the yield per unit area.

p = number of levels of factor i.

Intrinsically Nonlinear Model: This is when there is nonlinearity in parameters and linearity cannot be achieved through transformation.

The First Order Model;

$$y^{-1} = \beta_{11} + \beta_{01} x_1^{-1} + + \beta_{10} x_2^{-1} + + \beta_{00} (x_1 x_2)^{-1} \quad (3)$$

The Second Order Model;

$$y^{-1} = \beta_{11} + \beta_{01} x_1^{-1} + \beta_{10} x_2^{-1} + \beta_{02} x_2 x_1^{-1} + \beta_{20} x_1 x_2^{-1} + \beta_{00} (x_1 x_2)^{-1} \quad (4)$$

For details on properties of the nonlinear model [See Nduka, E. C. 1994, 1997].

### 3.1 Parameter Estimation Of A Nonlinear Function

When the normal equation is extremely difficult to solve, it may happen to have multiple solution to multiple stationary values of function of S(θ). It is said to have no closed form solution. Thereby the procedure of iterative method must be approached in order to manoeuvre the mathematical intractable problem (Bello O.A. 2014).

### 3.2 Nonlinearity of the Model.

Considering the parameter $\theta$, the estimate ($\hat{\theta}$) of $\theta$ is obtained by differentiating equation S($\hat{\theta}$) and equated to zero, resulting to p-1 normal equations. Also, considering the nonlinear model in (4). The sum of square error for our nonlinear model follows the definition;

$$S(\theta) = \sum_{i=1}^{n} [y_i - f(x_i, \theta_j)]^2 \quad (5)$$

$$\frac{x_1 x_2}{y} = \beta_{oo} + \beta_{10} x_1 + \beta_{01} x_2 + \beta_{20} x_2^2 + \beta_{02} x_1^2 + \beta_{11} x_1 x_2$$

Taking inverse of $y^{-1}$

$$y = [\beta_{11} + \beta_{01} x_1^{-1} + \beta_{10} x_2^{-1} + \beta_{20} x_1 x_2^{-1} + \beta_{02} x_2 x_1^{-1} + \beta_{00} (x_1 x_2)^{-1}]^{-1}$$

Let $\beta_j = \theta_j$

$$f(x, \theta) = [\theta_{11} + \theta_{01} x_1^{-1} + \theta_{10} x_2^{-1} + \theta_{20} x_1 x_2^{-1} + \theta_{02} x_2 x_1^{-1} + \theta_{00} (x_1 x_2)^{-1}]^{-1} \quad (6)$$

Recall the sum of square $S(\theta)=\sum_{i=1}^{n}(y-\hat{y})^2, E(y)=f(x,\theta)$ $given \quad E(\varepsilon)=0$ , thereby;

$$S(\theta)=\sum_{i=1}^{n}[y_i-(\theta_{11}+\theta_{01}x_1^{-1}+\theta_{10}x_2^{-1}+\theta_{20}x_1x_2^{-1}+\theta_{02}x_2x_1^{-1}+\theta_{00}(x_1x_2)^{-1})^{-1}] \tag{7}$$

The estimation of parameters $\theta_j$ is obtained by differentiating equation $S(\theta)$ for $\psi_{ij}$ equated to zero and solving for the $\theta_{s'}$ respectively.

$$S(\theta)=\sum_{i=1}^{n}[y_i-(\theta_{11}+\theta_{01}x_1^{-1}+\theta_{10}x_2^{-1}+\theta_{20}x_1x_2^{-1}+\theta_{02}x_2x_1^{-1}+\theta_{00}(x_1x_2)^{-1})^{-1}]\frac{\partial f(x,\theta)}{\partial \theta_j} \tag{8}$$

When the normal equation contains parameters that are depending on each other, we have a nonlinear problem which can be solved iteratively using computer software.

$$\underbrace{\begin{aligned}&\sum_{i=1}^{n}[y_i-f(x_i\theta)][\theta_{11}+\theta_{01}x_1^{-1}+\theta_{10}x_2^{-1}+\theta_{20}x_1x_2^{-1}+\theta_{02}x_2x_1^{-1}+\theta_{00}(x_1x_2)^{-1})^{-2}](x_1x_2)^{-1}=0\\&\sum_{i=1}^{n}[y_i-f(x_i,\theta)][\theta_{11}+\theta_{01}x_1^{-1}+\theta_{10}x_2^{-1}+\theta_{20}x_1x_2^{-1}+\theta_{02}x_2x_1^{-1}+\theta_{00}(x_1x_2)^{-1})^{-2}]x_1^{-1}=0\\&\sum_{i=1}^{n}[y_i-f(x_i,\theta)][\theta_{11}+\theta_{01}x_1^{-1}+\theta_{10}x_2^{-1}+\theta_{20}x_1x_2^{-1}+\theta_{02}x_2x_1^{-1}+\theta_{00}(x_1x_2)^{-1})^{-2}]x_2^{-1}=0\end{aligned}}_{The\,normal\,equation} \tag{9}$$

The normal equation in (9) is produced by the differentiation of $S(\theta)$ which involved model parameters that are nonlinear in relationship, the equation is mathematically intractable, cannot be solved analytically but could only be solved iteratively.

### 3.3 The Gauss Newton Method.

Choice of iterative algorithm is to minimize the sum of square error of the parameter estimates. The nonlinear least square surface is approximated based on the linear approximation, until no further improvement can be made, using the result of linear least squares in successive stages. At that point of no further improvement in decrease of sum of squared error, we say the steps have converged to a nonlinear least square solution. Unfortunately, one cannot always guarantee that such a least square solution will be achieved; in some cases the iteration can converge to a local minimum or 'pocket', and not the global least square solution but with good initial values, nonlinear least square algorithms will not perform well but if the initial estimates are far from the optimum point, it may perform well. The analyst must therefore provide intelligent, accurate starting values whenever employing iterative optimization algorithms.

In this work, we considered the SAS system search grid function to aid in selecting good initially value, also a self starter function in R-language can be use to generate initial parameters values. We also compare the performance of the two software in use. Furthermore, we shall seek to know if our iteration has generated a local minimum or global minimum estimates.For details on iterative algorithm methods, See (Analyzing Environmental Data W. W. Piegorsch and A. J. Bailer 2005 John Wiley & Sons, Ltd ISBN: 0-470-84836-7 (HB) page 40-46; Applied Linear Statistical Models Fifth Edition Michael H. Kutner, Emory University Christopher J. Ch 13; P g).

Using Gauss Newton method; let $\vartheta_o^o, \vartheta_2^o, ..., \vartheta_{p-1}^o$ be our stating values for parameters $\theta_0, \theta_1, ..., \theta_{p-1}$ through intelligent guess or default search grid in SAS or R-self starter function. The starting values s is supplied and we approximate the mean response $f(x,\theta)$ for n-case and using the first order of the Taylor series expansion of $f(x,\theta)$;

$$f(x,\theta) = f(x_{i,}\vartheta^o) + [\frac{\partial f(x_{i,}\theta)}{\partial \theta_k}]_{\theta=\vartheta}(\theta_k - \vartheta_k^o) \qquad (10)$$

By substitution; $y = f(x,\theta) + \varepsilon$

$$y = f(x_{i,}\vartheta^o) + [\frac{\partial f(x_{i,}\theta)}{\partial \theta_k}]_{\theta=\vartheta}(\theta_k - \vartheta_k^o) + \varepsilon \qquad (11)$$

For k parameters and n cases, i=1, 2... p

$$y_i - f(x_{i,}\vartheta^o) = \frac{\partial f(x_{1,}\theta)}{\partial \theta_o}\Big|_{\theta=\vartheta}(\theta_0^o - \vartheta_0^o) + \frac{\partial f(x_{2,}\theta)}{\partial \theta_1}\Big|_{\theta=\vartheta}(\theta_1^o - \vartheta_1^o) + \ldots + \frac{\partial f(x_{p,}\theta)}{\partial \theta_{p-1}}\Big|_{\theta=\vartheta}(\theta_{p-1}^o - \vartheta_{p-1}^o) \qquad (12)$$

For simplicity purpose, let $f_i^{(0)} = f(x_i, \vartheta^{(0)})$

and $\beta_k^{(0)} = (\theta_k - \vartheta_k^0)$ and $\Psi_{ik}^{(o)} = [\frac{\partial f(X_i, \theta)}{\delta \theta_k}]_{\theta=\vartheta^{(0)}}$

for ith observation for equation (12)

$$y_i - f_i^{(0)} = \sum_{k=0}^{p-1} \psi_{ik}^{(0)} \beta_k^{(o)} + \varepsilon_i$$

$$i = 1,2,\ldots,n \qquad (13)$$

$$y_i - f_i^o = \psi_{10}^{(o)}\beta_0^{(0)} + \psi_{11}^{(o)}\beta_1^{(0)} + \ldots, \psi_{1p-1}^{(o)}\beta_{p-1}^{(0)} + \varepsilon_i$$

For i=1,2,…,n

$$\begin{array}{lllll} y_i - f_i^o = & \psi_{10}^{(o)}\beta_0^{(0)} + \psi_{11}^{(o)}\beta_1^{(0)} + \ldots, & \psi_{1p-1}^{(o)}\beta_{p-1}^{(0)} & + \varepsilon_i \\ y_2 - f_2^o = & \psi_{20}^{(o)}\beta_0^{(0)} + \psi_{21}^{(o)}\beta_1^{(0)} + \ldots, & \psi_{1p-1}^{(o)}\beta_{p-1}^{(0)} & + \varepsilon_2 \\ \vdots & \vdots & \vdots & \vdots \\ y_n - f_n^o = & \psi_{n0}^{(o)}\beta_0^{(0)} + \psi_{n1}^{(o)}\beta_1^{(0)} + \ldots, & \psi_{np-1}^{(o)}\beta_{p-1}^{(0)} & + \varepsilon_n \end{array}$$

Writing the above in Matrix form;

$$\begin{bmatrix} y_i - f_1^o \\ \cdot \\ \cdot \\ \cdot \\ y_n - f_n^0 \end{bmatrix} = \begin{bmatrix} \Psi_{10}^{(o)} & \Psi_{11}^{(o)} & \ldots & \Psi_{1p-1}^{(o)} \\ \cdot & \cdot & \ldots & \\ \cdot & \cdot & \ldots & \\ \cdot & \cdot & & \\ \Psi_{n0}^{(o)} & \Psi_{n2}^{(o)} & \ldots & \Psi_{n,p-1}^{(o)} \end{bmatrix} \begin{bmatrix} \beta_0^{(0)} \\ \beta_1^{(0)} \\ \cdot \\ \cdot \\ \cdot \\ \beta_{p-1}^{(0)} \end{bmatrix} + \begin{bmatrix} \varepsilon_i \\ \varepsilon_2 \\ \cdot \\ \cdot \\ \cdot \\ \varepsilon_n \end{bmatrix} \qquad (14)$$

$$[Y - f^o]_{n\times1} = \psi_{n\times p-1}^0 \beta_{n\times1}^o + \varepsilon_{n\times1}$$

$Y - f^o$ is our column vector, $\psi_{n\times p-1}^0$ is the determinant matrix of known coefficients and $\varepsilon_{n\times1}$ is the disturbance.

To be able to make use of each result of the linear least squares in successive stages we use the least square estimates $\hat{\beta}_{(n\times1)}$ of $\beta$ written by;

$$\varepsilon = (y - f^0) - \psi^0\beta^0$$
$$\varepsilon^2 = [(y - f^0) - \psi^0\beta^0]^2$$
$$\varepsilon^2 = [(y - f^0) - \psi^0\beta^0][(y - f^0) - \psi^0\beta^0]^2$$
$$\partial\varepsilon^2 / \partial\beta = (y - f^o)^2 - 2(y - f^o)(\psi^o\beta^o) + (\psi^o\beta^o)^2$$
$$\psi^{o'}\psi^o\beta^o = \psi^o(y - f^o)$$

$$\hat{\beta}^o = (\psi^{o'}\psi^o)^{-1}\psi^o(y - f^o).$$

$\hat{\beta}^o$ is the least square estimates. The starting values are supplied for the parameters in $\psi$. The value of $\theta$ s that minimized the sum of square $S(\theta)$ is of interest, it's the point where we can be assured of a global minimum estimates.

$$SSE^{(o)} = \sum_{i=1}^{n}[y_i - f(x_i, \vartheta^{(o)})]^2$$

At the end of the first iteration we would have coefficient $\vartheta^{(1)}$ and the criterion measures as;

$$SSE^{(1)} = \sum_{i=1}^{n} [y_i - f(x_i, \vartheta^{(1)})]^2$$

The iteration should terminate where the SSE* becomes negligible or unchanging at point;

$$\left| \frac{(\theta_{i(p-1)} - \theta_{i,p})}{\theta_{ip}} \right| < \delta \quad i = 1,2,...,n \qquad (17)$$

### 3.4 The Gauss-Newton Application to IPM Second Order Model

Recall

$$y = [\beta_{11} + \beta_{01}x_1^{-1} + \beta_{10}x_2^{-1} + \beta_{20}x_1x_2^{-1} + \beta_{02}x_2x_1^{-1} + \beta_{00}(x_1x_2)^{-1}]^{-1}$$

and $\beta_j = \theta_j$, therefore from (13), we have;

$$y_i - f_i^0 = \sum_{k=0}^{p-1} \psi_{ik}^{(0)} \beta_k^{(o)} + \varepsilon_i$$ ; for $k = 0,1,2,3,4,5.$ and $i = 1,2,...,n$

$$y_i - f_1^0 = \psi_{i0}\beta_0 + \psi_{i1}^{(o)}\beta_1 + \psi_{i2}^{(o)}\beta_2 + \psi_{i3}^{(o)}\beta_3 + \psi_{i4}^{(o)}\beta_4 + \psi_{i5}^{(o)}\beta_5 + \varepsilon \qquad (18)$$

Recall $\Psi_{ik}^{(o)} = [\frac{\partial f(X_i,\theta)}{\delta\theta_k}]_{\theta_k=\vartheta_k^{(0)}}$, now for $k = 0,1,2,3,4,5.$

$$\Psi_{io}^{(o)} = [\frac{\partial f(X_i,\theta)}{\delta\theta_{00}}]_{\theta_k=\vartheta_k^{(0)}} = [(\theta_{11} + \theta_{01}x_1^{-1} + \theta_{10}x_2^{-1} + \theta_{20}x_1x_2^{-1} + \theta_{02}x_2x_1^{-1} + \theta_{00}(x_1x_2)^{-1})^{-1}](x_1x_2)^{-1}$$

$$\Psi_{i1}^{(o)} = [\frac{\partial f(X_i,\theta)}{\delta\theta_{01}}]_{\theta_1=\vartheta_1^{(0)}} = [(\theta_{11} + \theta_{01}x_1^{-1} + \theta_{10}x_2^{-1} + \theta_{20}x_1x_2^{-1} + \theta_{02}x_2x_1^{-1} + \theta_{00}(x_1x_2)^{-1})^{-1}]x_1^{-1}$$

$$\Psi_{i2}^{(o)} = [\frac{\partial f(X_i,\theta)}{\delta\theta_{02}}]_{\theta_2=\vartheta_2^{(0)}} = [(\theta_{11} + \theta_{01}x_1^{-1} + \theta_{10}x_2^{-1} + \theta_{20}x_1x_2^{-1} + \theta_{02}x_2x_1^{-1} + \theta_{00}(x_1x_2)^{-1})^{-1}]x_2x_1^{-1}$$

$$\Psi_{i3}^{(o)} = [\frac{\partial f(X_i,\theta)}{\delta\theta_{11}}]_{\theta_3=\vartheta_3^{(0)}} = 0$$

$$\Psi_{i4}^{(o)} = [\frac{\partial f(X_i,\theta)}{\delta\theta_{10}}]_{\theta_4=\vartheta_4^{(0)}} =$$
$$[(\theta_{11} + \theta_{01}x_1^{-1} + \theta_{10}x_2^{-1} + \theta_{20}x_1x_2^{-1} + \theta_{02}x_2x_1^{-1} + \theta_{00}(x_1x_2)^{-1})^{-1}]x_2^{-1}$$

$$\Psi_{i5}^{(o)} = [\frac{\partial f(X_i,\theta)}{\delta\theta_{20}}]_{\theta_5=\vartheta_5^{(0)}} = \quad (16.5)$$
$$= [(\theta_{11} + \theta_{01}x_1^{-1} + \theta_{10}x_2^{-1} + \theta_{20}x_1x_2^{-1} + \theta_{02}x_2x_1^{-1} + \theta_{00}(x_1x_2)^{-1})^{-1}]x_1x_2^{-1}$$

$$y_i - f_i^o = \psi^o{}_{i0}\beta_0{}^o + \psi_{i1}^{(o)}\beta_1{}^o + \psi_{i2}^{(o)}\beta_2^o + \psi_{i3}^{(o)}\beta_3^o + \psi_{i4}^{(o)}\beta_4^o + \psi_{i5}^{(o)}\beta_5 + \varepsilon_i$$

For I =1 ,2,…,n then;

$$[y_i - (\theta_{11} + \theta_{01}x_1^{-1} + \theta_{10}x_2^{-1} + \theta_{20}x_1x_2^{-1} + \theta_{02}x_2x_1^{-1} + \theta_{00}(x_1x_2)^{-1})^{-1}] = \psi_{i0}\beta_0 + \psi_{i1}^{(o)}\beta_1 + \psi_{i2}^{(o)}\beta_2 + \psi_{i3}^{(o)}\beta_3 + \psi_{i4}^{(o)}\beta_4 + \psi_{i5}^{(o)}\beta_5 + \varepsilon_i$$

$\psi_{i0}\beta_0 = \psi_{i0}\beta_{00}(x_1x_2)^{-1}, \psi_{i1}^{(o)}\beta_1 = \psi_{i1}^{(o)}\beta_{01}x_1^{-1},$

$\psi_{i2}^{(o)}\beta_2 = \psi_{i2}^{(o)}\beta_{10}x_2^{-1},$

$\psi_{i3}^{(o)}\beta_3 = \psi_{i3}^{(o)}\beta_{11}, \psi_{i4}^{(o)}\beta_4 = \psi_{i4}^{(o)}\beta_{10}x_1x_2^{-1},$

And $\psi_{i5}^{(o)}\beta_5 = \psi_{i5}^{(o)}\beta_{20}x_2x_1^{-1}.$

For k=0

$$\psi_{i0}\beta_0 = [(\theta_{11} + \theta_{01}x_1^{-1} + \theta_{10}x_2^{-1} + \theta_{20}x_1x_2^{-1} + \theta_{02}x_2x_1^{-1} + \theta_{00}(x_1x_2)^{-1})^{-2}]\beta_0$$

Let

$$\varphi = [(\theta_{11} + \theta_{01}x_1^{-1} + \theta_{10}x_2^{-1} + \theta_{20}x_1x_2^{-1} + \theta_{02}x_2x_1^{-1} + \theta_{00}(x_1x_2)^{-1}]$$

Thereby;

$$\psi_{i0}\beta_0 = \varphi^{-2}(x_1x_2)^{-1}\beta_{00}$$

The same for k=1, 2, 3, 4, and 5, then we have;

$\psi_{i1}^{(o)}\beta_1 = \varphi^{-2}x_1^{-1}\beta_{01},$

$\psi_{i2}^{(o)}\beta_2 = \varphi^{-2}x_2^{-1}\beta_{10},$

$\psi_{i3}^{(o)}\beta_3 = 0,$

$\psi_{i4}^{(o)}\beta_4 = \varphi^{-2}x_1x_2^{-1}\beta_{10}$ and

$\psi_{i5}^{(o)}\beta_5 = \varphi^{-2}x_2x_1^{-1}\beta_{20}$ respectively.

Recall (13)

$$y_i - f_i^0 = \sum_{k=0}^{p-1} \psi_{ik}^{(0)} \beta_k^{(o)} + \varepsilon_i \qquad i = 1,2,...,n$$

Making use of model components

$$y_i - \varphi^{-1} = \varphi^{-2}(x_{i1}x_{i2})^{-1}\beta_{00} + \varphi^{-2}x_{i1}^{-1}\beta_{01} + \varphi^{-2}x_{i2}^{-1}\beta_{10} + \varphi^{-2}x_{i1}x_{i2}^{-1}\beta_{10} + \varphi^{-2}x_{i2}x_{i1}^{-1}\beta_{20} + \varepsilon_i$$

Given n-observations, (19) is re-written in column-row matrix form;

$$\begin{bmatrix} y_1-\varphi^{-1} \\ y_2-\varphi^{-1} \\ . \\ . \\ . \\ y_n-\varphi^{-1} \end{bmatrix} = \begin{bmatrix} \varphi^{-2}(x_{11}x_{21})^{-1} & \varphi^{-2}x_{11}^{-1} & \varphi^{-2}x_{21}^{-1} & \varphi^{-2}x_{11}x_{21}^{-1} & \varphi^{-2}x_{21}x_{11}^{-1} \\ \varphi^{-2}(x_{12}x_{22})^{-1} & \varphi^{-2}x_{22}^{-1} & \varphi^{-2}x_{22}^{-1} & \varphi^{-2}x_{12}x_{22}^{-1} & \varphi^{-2}x_{22}x_{12}^{-1} \\ . & & . & . & . \\ . & . & & . & . \\ . & . & & . & . \\ \varphi^{-2}(x_{1n}x_{2n})^{-1} & \varphi^{-2}x_{2n}^{-1} & \varphi^{-2}x_{2n}^{-1} & \varphi^{-2}x_{1n}x_{2n}^{-1} & \varphi^{-2}x_{2n}x_{1n}^{-1} \end{bmatrix} \begin{bmatrix} \beta_{00} \\ \beta_{01} \\ \beta_{11} \\ \beta_{10} \\ \beta_{20} \end{bmatrix} + \begin{bmatrix} \varepsilon_1 \\ \varepsilon_2 \\ \vdots \\ \varepsilon_1 \end{bmatrix}$$

Equation (20) follows (14) thereby (15) is applied to solve for $\hat{\beta}_s$. The starting values for the $\theta s$ is supplied for the 1st iteration recalling $\beta_k^{(0)} = (\theta_k - \vartheta_k^0)$ and

$$\varphi = [(\theta_{11} + \theta_{01}x_1^{-1} + \theta_{10}x_2^{-1} + \theta_{20}x_1x_2^{-1} + \theta_{02}x_2x_1^{-1} + \theta_{00}(x_1x_2)^{-1}]$$

If $S(\theta)$ is increasing in successive iterations, new initial parameter value will be selected. Equation (17) is considered for convergence condition.

### 3.5 Softwares for Computer-Intensive Statistics.

The Computer-intensive statistics is just statistical methodology which makes use of a large amount of computer time though tedious but we have greater benefits of easily working with larger datasets, we can now use more realistic models than settling for linear model when relationship is not truly linear. Statistical software has revolutionized the way we approach data analysis. (For details see Brian D. Ripley, Professor of Applied Statistics University of Oxford: How Computing has Changed Statistics (and is Changing) ripley@stats.ox.ac.uk
http://www.stats.ox.ac.uk/ripley)

The iterative algorithm and bootstrap re sampling method are by nature computer-intensive. Many packages and programs exist for performing nonlinear optimizations, and in particular bootstrap nonlinear least square regression. We considered the SAS System and the R-programming language for this work. The SAS system is a popular set of industrial and educational use software tools, which allow you to access, manage, present, and analyze data. It runs on many different computer platforms and is designed to work similarly on different operating systems. While the R-programming language is available at CRAN: http://cran.r-project.org/web/packages/nlrwr/index.html. The codes used for computation in this study will be active upon loading nlrwr of function nls ( ) tools and nsltools. All the above softwares has brought to us a computer-intensive statistics which is capable of putting us over various barriers to many statistical method such as the ones applied in this work.
(For details see; Introduction to SAS 1http://help.unc.edu/statistical/applications/sas/introsasprog.html and R,; CRAN: http://cran.r-project.org/web/packages/nlrwr/index.html)
.

### 3.6 Bootstrap Re-Sampling Method

Bootstrapping methodology has become a recognised technique for dealing with data with extremely non-normal distribution. When faced with the problem of small sample sizes, the failure of normality assumption, non-linear relation of variables. They prove particularly useful where very limited sample data are available (Nancy Barker, 2002). When we do not have the knowledge of the distribution, the standard parametric techniques cannot be reliably executed, thereby an alternative is required. Bootstrapping, a data-based simulation method for assigning measures of accuracy to statistical estimates helps to tackle any of the above cases. Many conventional statistical methods of analysis make assumptions about normality but when these assumptions are violated, such methods may fail. Bootstrapping, a data-based simulation method, is a good alternative. Through the method of Bootstrap an estimator bias is reduced, the method also gives the distribution of the statistic under consideration [(See Nancy Barker, 2002].

### 3.7 Gauss Newton-Bootstrap Re-Sampling Method.

Considering the observation $\mathbf{Y}_1, \mathbf{Y}_2, \mathbf{Y}_{3,...,} \mathbf{Y}_n$ b e i.i.d from an experimental design with unknown distribution.

$$\mathbf{Z}' = [Y_i, X_{1i}, ..., X_{ki}]_{\text{ i=1,2,...,n}} \qquad (21)$$

the observation in equation[above] can be sampled B times with replacement to have bootstrap samples $\mathbf{Z}'_1, \mathbf{Z}'_2, ..., \mathbf{Z}'_B$ through the use of computer intensive bootstrap re sampling method. Bootstrap observation $\mathbf{Z}'^*_\mathbf{i}$ is produced.

$$\mathbf{Z}'^*_\mathbf{i} = [Y_i, X_{1i}, ..., X_{ki}] \quad _{\text{i=1,2,...,B}} \qquad (22)$$

Where $X_1, ..., X_k$ is treated as fixed because the data of interest in this research work is derived from an experimental design. Having the $X_s$ fixed, to compute the nonlinear estimates for bootstrap re sampled samples, the Gauss- Newton Method is submerged into the Bootstrap re sampling method at this point. Recall (14) and (20) for i-observation per each $\mathbf{B}'^*_{i^*}$ cases where i=1,2,...,n and $i^* = 1,2,...,1000$ i.e $\mathbf{B}$ =1000; starting values is supplied to obtained the bootstrapped estimates $\theta'^*_{bi^*}$ .

$$\overbrace{\begin{array}{l} \theta'^*_{b1^*} = [\theta^*_{b0}, \theta^*_{b2}, ..., \theta^*_{bn}] \\ \theta'^*_{b2^*} = [\theta^*_{b0}, \theta^*_{b2}, ..., \theta^*_{bn}] \\ . \quad . \quad . \quad . \\ . \quad . \quad . \quad . \\ . \quad . \quad . \quad . \\ \theta'^*_{\mathbf{b100}} = [\theta^*_{b0}, \theta^*_{b2}, ..., \theta^*_{bn}] \end{array}}^{\text{B}_i \text{ cases}} \qquad (23)$$

The bootstrapped estimates $\theta'^*_{bi^*}$ is obtained at the point where SSE* is minimum. See section (3.3). The estimates in (23) is used to estimate the fitted values and residuals for each observation in $\mathbf{Z}'^*_\mathbf{i}$.

$$\hat{Y}_i = f(x_i, \theta_i) \qquad (24)$$

Where $\mathbf{E}_i = \mathbf{Y_i} - \hat{\mathbf{Y}}_\mathbf{i}$ are the residuals. The next procedure is to sample residual n-times for each bootstrap samples B, and each residuals of the bootstrap samples $\mathbf{E}'^*_{b1^*} = [\varepsilon^*_{b0}, \varepsilon^*_{b2}, ..., \varepsilon^*_{bn}]$ is attached to the deterministic component of the model $f(x_i, \theta_i)$ to obtain the bootstrapped observations.

$$\mathbf{Y}_{bi}{}^* = [Y^*_{b1}, Y^*_{b2}, ..., Y^*_{bn}], \quad _{\text{i=1,2,...,B}} \qquad (25)$$

Where $\mathbf{Y}_{bi}{}^* = \hat{Y}_{bi} + E_{bi}$.

This is followed by regressing each bootstrapped $\mathbf{Y}_{bi}{}^*$ on the fixed $\mathbf{X}_i s$ to obtained the bootstrapped regression coefficients $\theta'^*_{bi^*}$ It worth noting that the functional form of our model in this work is a non-linear approximation (recall section 3.2), thereby, all estimation is carried out iteratively , which is also computer demanding. The re sampled bootstrap coefficients $\theta'^*_{bi^*}$ is now used to construct bootstrapped standard error, confidence interval and other graphical techniques to be used in drawing our inference.

### 3.8 Bootstrapped Standard Error (Std Error)

The bootstrapped values $\theta'^*_{bi^*}$ for ith observation and B cases are used to estimate the standard error calculated by:

$$Se_b(\theta^*) = \sqrt{\frac{1}{B}\sum_{i=1}^{B}(\hat{\theta}_i^* - \hat{\theta})^2}$$

### 3.9 Bootstrapped Confidence Interval (C.I)

The bootstrapped estimates of the Std error and normality assumption arbitrary will be use to construct an interval in form;

$$[\hat{\theta}^* \pm Z_{i-\frac{\alpha}{2}} Se(\hat{\theta}^*)] \qquad (27)$$

Where $Z\alpha$ denote the $\alpha$ quantile of standard normal distribution. This is $(1-\alpha)C.I$ for $\hat{\theta}$.

## 4 DATA PRESENTATION, CODES AND RESULTS

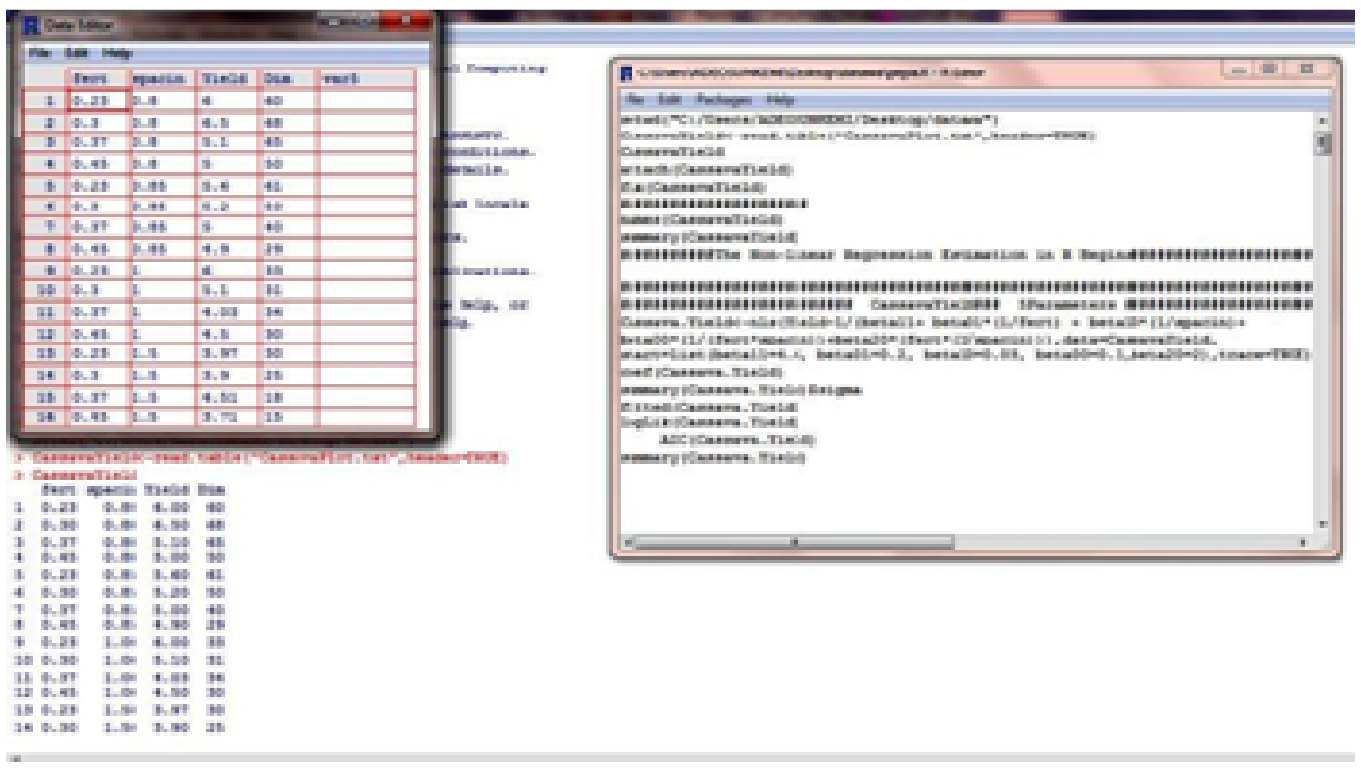

Figure 1.

### 4.1 Graphical Analyses

#### Prediction Plots

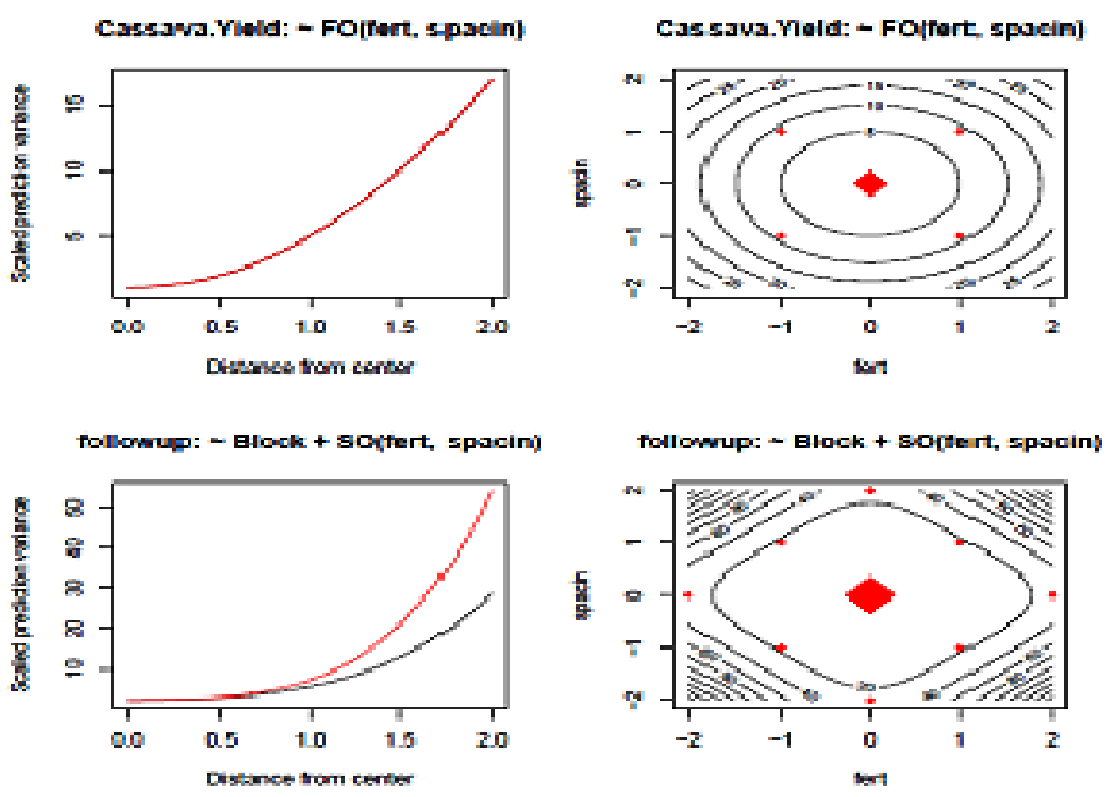

Figure 2.

Variance Function (varfcn) Plots for Design Properties Check:

The helpful tool in 'rsm' R-package is the varfcn function, which helps to see the variance of the predictions we will obtain making use of scaled variance defined by $N/\delta^2 \, Var[\hat{y}(x)]$ and $\hat{y}(x)$ is the predicted value at the design point x . The varfcn function can also help to verify if any two of the cube blocks plus the axis block is sufficient to estimate a second-order response surface. The y-axis is the scaled prediction variance and the x-axis measured as distance from the centre. For details see Response-surface illustration,[20].

Inspecting the plots in fig 2 above, the variance function plot for properties check for the 1st [FO (fert, spacin)] and 2nd [SO (fert, spacin)] order designs respectively; variance will increases as we go farther away from the centre, this means making estimation at or closer to the centre will give accurate predictions. The closer the prediction from the centre, the more accurate the predicted values. In fig 2 of 'FO (fert, spacin)] ', the predicted variance dotted line is exactly on the red line representing the center of the surface from all direction, this assures the same variance in all direction and the contour plots circular formation shows rotatability properties is achieved with the chosen designs. The second plot [SO (fert, spacin)] in fig 2 shows the black line is not too far from the centre (red line) and it shows a little less perfect circle for the augmented second order design. With this we can at least augment the design to fit a second-order model close enough to fulfil the uniform precision and rotatability conditions. This implies we can go ahead and fit our model. See appendix III for the constructed designs.

### 4.2 EDA Plots for Cassava Plantation Experimental Data.

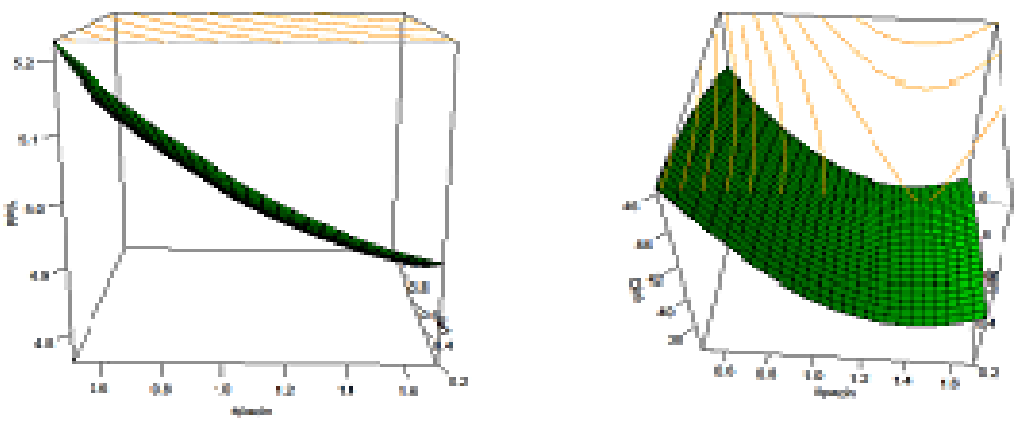

Figure 3.

The exploratory data analysis(EDA) of our dataset in fig 3 suggest a nonlinear function as the relationship describing the plant-yield phenomenon, having fertilizer(fert) and crop spacing(spacing) as the explanatory variables.

| Parameters | beta11 | beta01 | beta10 | beta00 | beta20 |
|---|---|---|---|---|---|
| Bootstrap Estimates | 0.3560391 | -0.0092282 | -0.220062 | 0.011455 | 0.20223 |
| Bootstrap C.I | | | | | |
| 2.5% | 0.15002602 | -0.06944452 | -0.54175 | -0.05147 | -0.24210 |
| 97.5% | 0.55620729 | 0.05038302 | 0.09993 | 0.07746 | 0.612189 |

Table1

| | SAS System | | R-nls | | | |
|---|---|---|---|---|---|---|
| Iteration No. | 9 | | 9 | | Confidence Interval | |
| Parameters | Estimates | Approx. Std Error | Estimates | Approx.Std Error | 2.5% | 97.5% |
| $\beta_{11}$ | 0.3526 | 0.1271 | 0.3494492 | 0.121360 | 0.091022 | 0.632016 |
| $\beta_{01}$ | -0.0667 | 0.0843 | -0.0084451 | 0.035495 | -0.086046 | 0.070524 |
| $\beta_{10}$ | -0.2112 | 0.2014 | -0.216099 | 0.194589 | -0.656282 | 0.2130867 |
| $\beta_{00}$ | 0.0407 | 0.0565 | 0.0113127 | 0.039490 | -0.073257 | 0.099431 |
| $\beta_{20}$ | 0.1870 | 0.2661 | 0.1978421 | 0.259470 | -0.385833 | 0.793252 |

### 4.3 Discussion Of Results

Model Adequacy

The underlying assumptions of normality, independent, constant variance and homogeneity of the model error were used to verify the adequacy of our function used to describe the dataset.

The QQ-plot shows a good agreement between the distribution of the model residual and that of the standard normal distribution. This shows the adequacy and fitness of the model to the data; seeing the pattern is linear and the points falling within straight line of the plot i.e. points are closer to the straight line (See appendix I). The plot does not suggest any serious departure from our model assumptions of normality. The plot of the residual against the predictors, appendix II, show a non-systematic or random pattern in the model residual plots, therefore variance homogeneity assumption is intact. The various plots as presented in the appendix gave a basis to submit that error variance of our function is normal, independent, constant and there is consistency of variance of error term.

The choice of starting value is vital with Gauss-Newton method in order to achieve quick convergence and a global minimum estimate. An approximated estimate from linear least square method and grid search values were used as starting values for the parameters and the final result (estimate values, SSE of the model) for both R and SAS were found to be in agreement. The SAS System Iteration converges after 9 iterations with the model sum of square error (SSE) 2.3600. The R-nls() Iteration also converges after 9 iterations with sum of square error (SSE) 2.247653[See appendix IV and V]. In both cases the five parameters values falls within our confidence interval range. Also to see if we are able to achieve a global minimum estimate, we re-introduced the obtained estimates values as the starting values into our nonlinear regression model and the same estimate values were obtained with their sum of square error unchanged, a global minimum estimates were achieved, see appendix VI.

## 5 Conclusion

The computer intensive method-bootstrap re sampling method gives an access over many statistical barriers involving fewer observation or sample size, such as demonstrated in this work. It has helped us in this work to evaluate the precision of the obtained estimates and we can confidently present the model with reliable estimates that relate the non-linear relationship between cassava yields to various amount of fertilizer applied and crop spacing. The accompanied model is adequate and sufficient for the non-linear response surface.

### 5.1 Appendices

*Normally distributed errors *Variance consistency

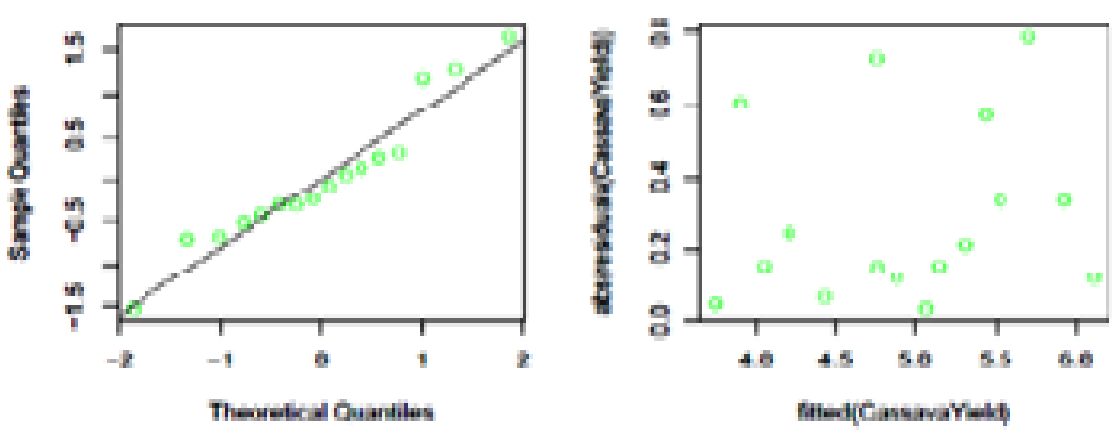

Appendix I

*variance homogeneity

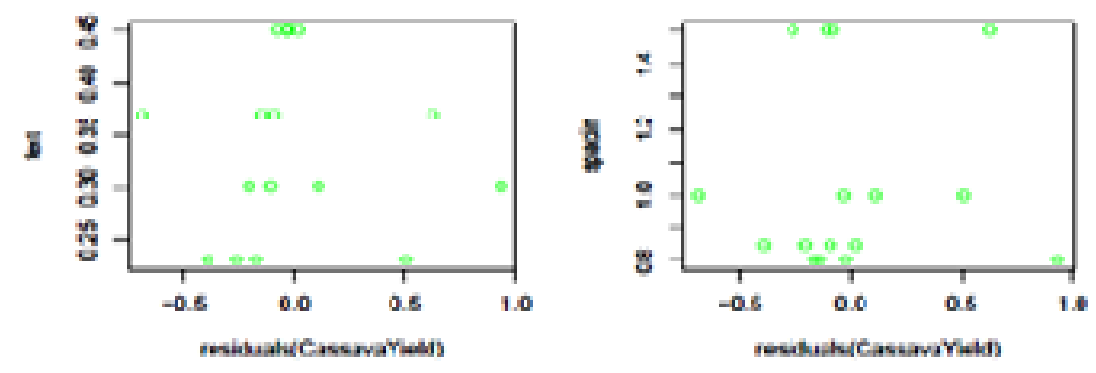

Appendix II

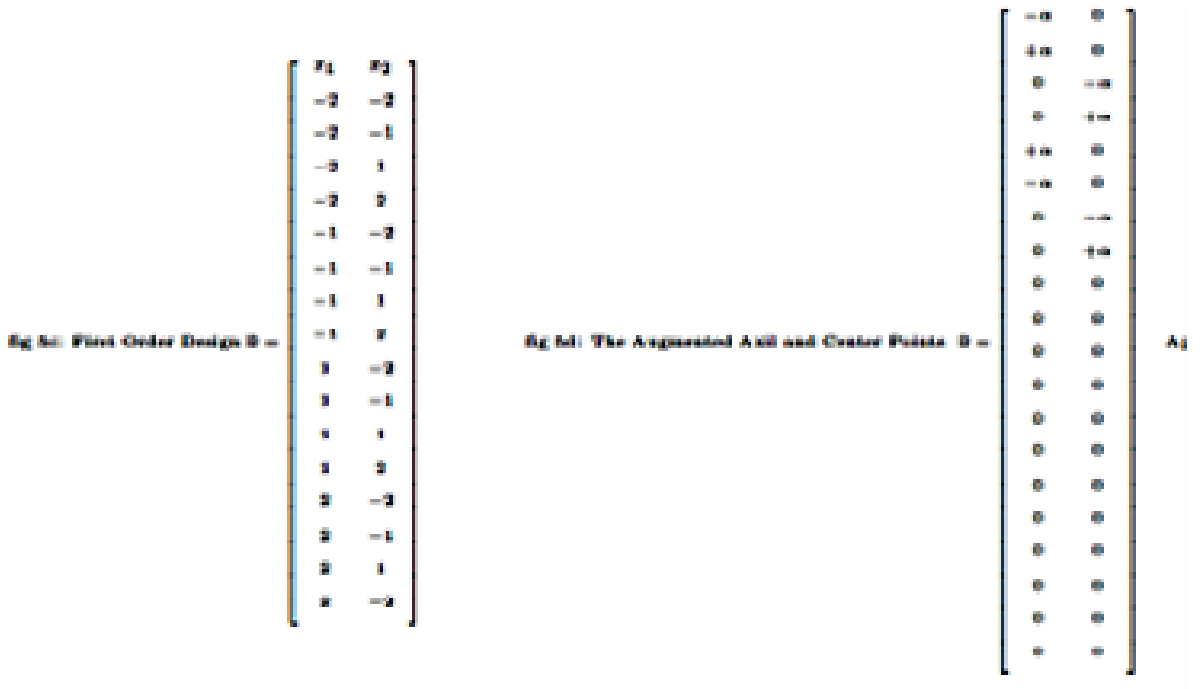

Appendix III

**SAS Results**

Appendix V

16:28 Wednesday, October 1, 2014 pt 12

The NLIN Procedure
Dependent Variable CassavaYield
Method: Gauss-Newton

Iterative Phase

| Iter | beta11 | beta01 | beta10 | beta00 | beta20 | Sum of Squares |
|---|---|---|---|---|---|---|
| 0 | 10.0000 | 0.2000 | 0.0300 | 0.2000 | 0.2000 | 386.8 |
| 1 | 6.9126 | 0.6646 | -0.0759 | -0.9165 | 1.3215 | 375.4 |
| 2 | 5.0498 | -0.2366 | -3.8504 | 0.3428 | 3.4081 | 340.3 |
| 3 | 1.7520 | -0.0194 | -0.9672 | 0.0236 | 0.9121 | 265.4 |
| 4 | 0.8748 | -0.0252 | -0.5634 | 0.0342 | 0.5089 | 141.2 |
| 5 | 0.4989 | -0.0117 | -0.3048 | 0.0157 | 0.2817 | 38.7004 |
| 6 | 0.3111 | -0.00773 | -0.1934 | 0.0104 | 0.1765 | 8.8185 |
| 7 | 0.3451 | -0.00839 | -0.2137 | 0.0112 | 0.1955 | 2.4333 |
| 8 | 0.3494 | -0.00845 | -0.2161 | 0.0113 | 0.1978 | 2.3600 |
| 9 | 0.3495 | -0.00845 | -0.2161 | 0.0113 | 0.1978 | 2.3600 |

Appendix VI

The NLIN Procedure
Dependent Variable CassavaDia
Method: Gauss-Newton

Iterative Phase

| Iter | beta11 | beta01 | beta10 | beta00 | beta02 | beta20 | Sum of Squares |
|---|---|---|---|---|---|---|---|
| 0 | 0.1705 | -0.00771 | -0.1719 | 0.0182 | -0.00846 | 0.0831 | 309.0 |
| 1 | 0.1708 | -0.00791 | -0.1721 | 0.0184 | -0.00841 | 0.0831 | 309.0 |
| 2 | 0.1705 | -0.00770 | -0.1719 | 0.0182 | -0.00847 | 0.0831 | 309.0 |
| 3 | 0.1705 | -0.00772 | -0.1719 | 0.0182 | -0.00846 | 0.0831 | 309.0 |
| 4 | 0.1705 | -0.00771 | -0.1719 | 0.0182 | -0.00846 | 0.0831 | 309.0 |

NOTE: Convergence criterion met.